\documentclass{article}

\usepackage{amssymb,amsfonts,amsmath,stmaryrd}
\usepackage{cite,enumerate,float,indentfirst}
\usepackage{color}

\def\be{\begin{eqnarray}}
\def\ee{\end{eqnarray}}
\def\nn{\nonumber}

\def\p{\partial}

\newcommand{\beq}{\begin{equation}}
\newcommand{\eeq}{\end{equation}}
\newcommand{\beqa}{\begin{eqnarray}}
\newcommand{\eeqa}{\end{eqnarray}}

\definecolor{red}{rgb}{1,0,0}
\definecolor{orange}{rgb}{1,0.5,0}
\definecolor{violet}{rgb}{0.7,0,1}

\def\sn{{\rm sn}}

\hoffset= - 1.0in         

\begin{document}

\title{\vspace{1.5cm}\bf
pq-Duality: a set of simple examples
}

\author{
{Z. Zakirova$^{a,b}$}\footnote{zolya\_zakirova@mail.ru},\ \
{V. Lunev$^a$}\footnote{vladislavlunev138@gmail.com}\ \ and\
{ N. Beloborodov$^a$}\footnote{nichegoloch567@gmail.com}
}

\date{ }

\maketitle

\vspace{.2cm}

\begin{center}
$^a$ {\small {\it Kazan State Power Engineering University, Kazan, Russia}}\\
$^b$ {\small {\it Institute for Information Transmission Problems, Moscow 127994, Russia}}\\
\end{center}

\vspace{.1cm}

\begin{abstract}
A series of two-particle examples of the Ruijsenaars pq-duality is considered in detail, the dual Hamiltonians are constructed. Of special interest is the case of the sinh-Gordon model.
\end{abstract}

\bigskip

\section{Introduction
\label{intro}}

The notion of pq-duality was introduced by S. Ruijsenaars \cite{Rui} in description of many-body systems of the Calogero-Ruijsenaars family. It was later exploited in Seiberg-Witten theory for constructing the double-elliptic systems describing the low energy limit of the $6d$ SUSY gauge theory with adjoint matter hypermultiplet \cite{BMMM,MM}. In fact, the double-elliptic systems were constructed only in the simplest case of the system with one degree of freedom, and the question of constructing the $N$-body system remains open so far.

Later, in \cite{MM2}, it was formulated a more general framework to deal with the pq-duality, which allows one to treat it with any integrable $N$-body system. However, only a restricted set of examples of dual systems has been known so far. In this short note, we are going to extend this set. In particular, starting as a warm-up from the well-known example of the trigonometric Ruijsenaars system, where the duality was originally established by S. Ruijsenaars \cite{Rui} (though in an absolutely different way), we further study the case of the two-particle Toda system, non-periodic (Liouville theory) and periodic (sinh-Gordon theory). The later case of much importance, since we do not know the Hamiltonians dual to the elliptic Calogero or Ruijsenaars models so far, and, as we demonstrate here, the periodic Toda enjoys a hidden ellipticity, which does not come as a surprise since it is obtained from the elliptic Calogero system by the Inozemtsev limit \cite{Ino}. This hidden ellipticity shows up in the elliptic dual Hamiltonian, with the elliptic modulus depending on the coordinate. On the other hand, the periodic Toda is a simpler system than the elliptic Calogero model, hence, it will be probably easier to obtain the dual Hamiltonians also in the case with the number of particles $N>2$.

We consider here only the classical duality, though the quantum pq-duality is even more natural, and one would better start with the quantum case coming to the classical pq-duality in the quasiclassical limit. Still, the classical case by itself is of the same importance.

\section{Duality transformation in the systems with one degree of freedom}

Let us explain what is the pq-duality. Consider an integrable system with a property $H_i(p,q;0)=H_i^{(0)}(p)$. Duality is defined to be a map from a set of Hamiltonians $H_i(p,q|g)$ where $g$ is a coupling constant to a set of dual Hamiltonians $H^D_i(p,q|g)$ given by the following anti-canonical change of variables $\{p_i,q_i\}\to\{P_i,Q_i\}$:
\be\label{duality}
\left\{
\begin{array}{rcl}
H_i(p_i,q_i|g)&=&H_i^{(0)}(Q_i)\cr
&&\cr
\sum_idp_i\wedge dq_i&=&-\sum_idP_i\wedge dQ_i
\end{array}
\right.
\ee
This definition means that the Hamiltonians $H_i^{(0)}(Q_i)$ describe a free system.
There is a freedom in this definition, e.g., one can add to $P_i$ arbitrary functions of $Q_i$.

Then, a set of dual Hamiltonians is defined to be
\be\label{dualH}
H_i^D(P_i,Q_i|g)&=&H_i^{D,(0)}(q_i)
\ee
It is not much of importance how to choose $H_i^{D,(0)}$ since a function of Hamiltonians is still a Hamiltonian, however, one has to choose it in such a way that $H_i^D$ is a well-defined (single-valued) function on the symplectic manifold.

If the dual Hamiltonians can be chosen coinciding with the original ones, the system is called self-dual. The requirement of {\it anti}-canonical transformation $\{p_i,q_i\}\to\{P_i,Q_i\}$ is because, at $g=0$, the free system Hamiltonian $H_i^{(0)}(p_i)$ is equal to $H_i^{(0)}(Q_i)$, i.e. $p_i=Q_i$. One could certainly define the new coordinates in the canonical way, interchanging $Q_i$ and $P_i$, but then the Ruijsenaars model would became non-self-dual.

As a simple example, consider the Hamiltonian (this is called rational Calogero system)
\be\label{1}
H(p,q;g)={p^2\over 2}+{g\over 2q^2}
\ee
and
\be\label{2}
H_0(p)={p^2\over 2}
\ee
Now introduce new variables $P$, $Q$ such that:
\be\label{3}
H(p,q;g)=H_0(Q)
\ee

The transformation $(p,q)\longrightarrow (P,Q)$ is defined to be anti-canonical.  Thus, one has the two conditions for the two new variables $P$ and $Q$. In fact, the anti-canonical transformation is ambiguously defined: one can add to $P$ any arbitrary function of $Q$.

In these new variables, one can introduce a new free Hamiltonian $H_0^D$ depending only on $q$ and define
$H^D$ called dual Hamiltonian:
\be\label{5}
H^D(P,Q;g):=H_0^D(q)
\ee
For instance, for the Hamiltonian (\ref{1}) and equation (\ref{3}), the anti-canonical transformation with the simplest choice of the variable $P$ gives
\be\label{6}
Q^2=p^2+{g\over q^2}\nn\\
P^2=q^2-{g\over Q^2}
\ee
Indeed, the Poisson bracket
\be\label{7}
\{P,Q\}={\p P\over \p p}\cdot {\p Q\over \p q}-{\p Q\over \p p}\cdot {\p P\over \p q}=
-{g^2\over (p^2q^2+g)^2}-{p^2q^2q(2g+p^2q^2)\over(p^2q^2+g)^2}=-1
\ee
how it should be for an anti-canonical transformation. Now we choose $H_0^D(q)={q^2\over 2}$ and observe that
\be
H^D(P,Q)={P^2\over 2}+{g\over 2Q^2}=H(P,Q)
\ee
Hence, we have a self-dual Hamiltonian: the system is self-dual.

 The simplest way to deal with the anti-canonical transformation is to start with it in the form
\be\label{8}
dP\wedge dQ=-dp\wedge dq
\ee
Indeed, let us choose $H^D_0=H_0$ from the very beginning in our example. Then, from (\ref{3}),
\be\label{9}
{\p H_0(Q)\over \p Q}dQ={\p H(p,q;g)\over\p p}dp+{\p H(p,q;g)\over\p q}dq
\ee
and, from (\ref{5}),
\be\label{10}
{\p H_0^D(q)\over \p q}dq={\p H^D(P,Q;g)\over\p P}dP+{\p H^D(P,Q;g)\over\p Q}dQ
\ee
Multiplying the left hand side of (\ref{9}) with the right hand side of (\ref{10}) and vice versa, and using that $dQ\wedge dQ=dP\wedge dP=0$, one obtains
\be\label{11}
{\p H_0(Q)\over \p Q}\cdot {\p H^D(P,Q;g)\over \p P}={\p H_0^D(q)\over \p q}\cdot {\p H(p,q;g)\over \p p}
\ee
This equation, along with (\ref{3}) and (\ref{5}), provides 3 equations for 3 unknown variables $P$, $Q$ and $H^D(P,Q;q)$. However, any solution to these equations admits adding an arbitrary function of $Q$ to $H^D$.

 \section{The trigonometric Ruijsenaars system}

As a more non-trivial example, consider the hyperbolic two-particle Ruijsenaars systems. From now on, we work in the center of mass with the interaction Hamiltonians depending only on differences of coordinates. Hence, in the two-particle case, we choose $p=p_1=-p_2$ and $q=q_1-q_2$. The Ruijsenaars Hamiltonian in this case is
\be\label{13.5}
H={\sinh(q+g)\over\sinh q}e^p+{\sinh(q-g)\over\sinh q}e^{-p},
\quad
H_0^D(q)=2\cosh q
\ee
In this case,  $H(p,q;0)=H_0(p)=e^p+e^{-p}$, then $H_0(Q)=e^Q+e^{-Q}=2\cosh Q$. From (\ref{3}) one obtains
$$
2\cosh Q={\sinh(q+g)\over\sinh q}e^p+{\sinh(q-g)\over\sinh q}e^{-p}
$$
Write the equation  (\ref{11}) taking into account $H^D(P,Q;g)=2\cosh q$:
\be\label{14}
{\sinh Q}\cdot {\p H^D(P,Q;g)\over \p P}=\sinh q\cdot {\p H(p,q;g)\over \p p}
\ee
Note that
$$
\Big({\p H\over \p p}\Big)^{2}-H^{2}=4\cdot{\sinh(q+g)\sinh(q-g)\over \sinh^2q}
$$
Then equation  (\ref{14}) is
$$
{\sinh Q}\cdot {\p H^D\over \p P}=\sqrt {H^2\sinh^2q-4\sinh(q+g)\sinh(q-g)}
$$
with a properly chosen branch of the square root.

This equation, after transformations, can be written as an integral
$$
\int^{H^D}{dz\over\sqrt { z^2-k^2 }}=P,
\quad
k^2=4\cdot{\sinh^2Q-\sinh^2g\over \sinh^2{Q}}
$$
After performing the integration, one finds
\be\label{P}
P=\ln | H^D+\sqrt{(H^D)^2-k^2}|
\ee
and one can freely add an arbitrary function of $Q$ to $H^D$.
One can obtain from (\ref{P})
$$
2H^D=e^P+k^2{e^{-P}}
$$
and, rewriting $k^2$ as
$$
k^2=4\cdot{\sinh(Q+g)\sinh(Q-g)\over \sinh^2{Q}}
$$
and shifting $P\to P+\ln{2 \sinh(Q+g)\over \sinh Q}$, one finally obtains
\be\label{15}
H^D={\sinh(Q+g)\over\sinh Q}e^P+{\sinh(Q-g)\over\sinh Q}e^{-P}
\ee
Thus, from formulas  (\ref{13.5}) and  (\ref{15}), we reproduce the celebrated result due to S. Ruijsenaars \cite{Rui} that the hyperbolic Ruijsenaars system is self-dual. Taking into account $H^D(P,Q;g)=2\cosh q$, one gets that
$$
2\cosh q={\sinh(Q+g)\over\sinh Q}e^P+{\sinh(Q-g)\over\sinh Q}e^{-P}
$$

\section{The non-periodic Toda system}

Now we consider a new example of the dual systems, the non-periodic Toda system, which is given, in the two-particle case, by
\be
H(p,q;g)={p^2\over 2}+ge^q,
\quad
H_0^D(q)=e^q
\ee
When $g=0$, the free system is $H(p,q;0)=H_0(p)={p^2\over 2}$, i.e. $H_0(Q)={Q^2\over 2}$. Then, from (\ref{3}),
\be\label{12}
Q^2={p^2}+2ge^q
\ee
In accordance with formula (\ref{5}), the dual Hamiltonian $ H^D(P,Q;g)=e^q$, and, from  (\ref{11}), one obtains
$$
{Q}\cdot {\p H^D(P,Q;g)\over \p P}={pe^q}
$$
 or,
$$
{Q}\cdot {\p H^D\over \p P}=H^D\cdot\sqrt{Q^2-2gH^D}
$$
again with a properly chosen branch of the root.

\bigskip

\noindent
Integrating this equation with a properly chosen arbitrary function of $Q$, one obtains
\be
P=\ln{{{Q-\sqrt{Q^2-2gH^D}}\over Q+\sqrt{Q^2-2gH^D}}}
\ee
therefore, the dual Hamiltonian is
\be
H^D={Q^2 \over 2g\cosh^2(P/2)}.
\ee
The variable $P$ in terms of the $p,q$ variables is
\be\label{13}
P=\ln{{{{\sqrt{p^2+2ge^q}-p}}\over{\sqrt{p^2+2ge^q}+p}}}
\ee
Using formulas (\ref{12}), (\ref{13}), one can check the Poisson bracket:
$$
\{P,Q\}=-{2ge^q\over p^2+2ge^q}+{gpe^q\over p^2+2ge^q}\Big ({1\over{\sqrt{p^2+2ge^q}+p}}-{1\over{\sqrt{p^2+2ge^q}+p}}\Big)=
$$
$$
=-{2ge^q+p^2\over p^2+2ge^q}=-1
$$

\section{The periodic Toda system}

At last, we consider the most interesting example: that of the periodic Toda chain. In the two-particle case, the Hamiltonian is
\be\label{pT}
H(p,q;g)={p^2\over 2}+g^2\cosh^2q
\quad
H_0^D(q)=\cosh q
\ee
therefore,  $H(p,q;0)=H_0(p)={p^2\over 2}$, from here $H_0(Q)={Q^2\over 2}$, and from (\ref{3}),
\be
Q^2={p^2}+2g^2\cosh^2{q}.
\ee
The  equation  (\ref{11}) is of the form
$$
{Q}\cdot {\p H^D(P,Q;g)\over \p P}=p\sinh q
$$
 or,
$$
{Q}\cdot {\p H^D\over \p P}=\sqrt{Q^2-2g^2(H^D)^2}\cdot\sqrt{(H^D)^2-1}
$$
Here $\cosh q=H^D$, and one has to be careful about the {\it two} branches of the roots. One obtains integral
$$
\int^{H^D}{{dz}\over{\sqrt{Q^2-2g^2z^2}\cdot\sqrt{z^2-1}}}={P\over Q}
$$
It can be rewritten in the form
$$
\int^{H^D}{{dz}\over{\sqrt{1-k^2z^2}\cdot\sqrt{1-z^2}}}={iP},
\quad
k^2={2g^2\over Q^2}.
$$
After changing the variable $H^D=\sin t$, one obtain integral
$$
\int^{\arcsin H^D}_0{{dt}\over{\sqrt{1-k^2 \sin^2 t}}}={iP}
$$
which is an elliptic integral, and we have fixed a free function of $Q$ by choosing the lower limit of the integral. Now, it is easy to obtain the formula for the dual Hamiltonian $H^D$ in the form of the Jacobi function $\sn (x|k)$, where $k$ is the elliptic modulus \cite{BE}:
$$
H^D=\sn \Big(iP\Big|\sqrt{2}gQ^{-1}\Big)
$$
A surprising feature of this dual Hamiltonian is emergence of the elliptic function though the original Hamiltonian (\ref{pT}) has not contain any ellipticity. Moreover, the elliptic modulus in $H^D$ is dynamical, i.e. depends on the coordinate. The reason is that the periodic Toda chain is obtained by the Inozemtsev limit from the elliptic Calogero model \cite{Ino}, and the Hamiltonian dual to the elliptic Calogero one also depends on the dynamical elliptic modulus \cite{BMMM}.

Let us stress once more that this example is of great importance, since it catches the main features of the elliptic Calogero model (the dual Hamiltonian is the Jacobi function with a dynamical elliptic modulus \cite{BMMM}), but is much simpler: the original Hamiltonian is not elliptic. This opens a possibility of constructing dual Hamiltonians possessing main features of the elliptic Calogero case at the number of particles $N>2$.

\section*{Acknowledgements}

We are grateful to A. Mironov for valuable discussions. This work was supported by the Russian Science Foundation (Grant No.23-41-00049).

\end{document}